\documentclass[twocolumn,secnumarabic,amssymb, nobibnotes, aps, prd,superscriptaddress]{revtex4-2}
\usepackage{graphicx}

\usepackage{filecontents}

\begin{document}

\title{Stabilization of magnetic bubbles in [Ni/Co]$_{n}$ multilayers on an oxygen-reconstructed
Nb(110) surface via an ultra-thin Cu interlayer}%

\author{Ahmad Dibajeh}%
\affiliation{Zernike Institute for Advanced Materials, University of Groningen, 9747 AG Groningen, The Netherlands}
\author{Cameron W. Johnson}
\affiliation{Molecular Foundry, Lawrence Berkeley National Laboratory, Berkeley, California 94720, U.S.A.}
\author{Andreas K. Schmid}
\affiliation{Molecular Foundry, Lawrence Berkeley National Laboratory, Berkeley, California 94720, U.S.A.}
\author{Roberto Lo Conte}
\email[]{r.lo.conte@rug.nl}
\affiliation{Zernike Institute for Advanced Materials, University of Groningen, 9747 AG Groningen, The Netherlands}
\affiliation{Institute of Nanostructure and Solid State Physics, University of Hamburg, 20355 Hamburg, Germany}

\date{\today}%

\begin{abstract}
Magnetic thin films hosting topological spin textures, such as magnetic skyrmions, hold high potential for breakthroughs in the field of spintronics, due to good scalability and energy efficiency. Novel computational architectures such as memory-in-logic devices rely on material platforms able to host those topological spin textures. Furthermore, recently proposed designs of novel quantum information technologies are based on heterostructures where topological spin textures are in direct proximity to a superconducting layer. 
Here, we demonstrate the stabilization of out-of-plane magnetic bubbles in highly ordered [Ni/Co]$_{n}$ multilayers on a Nb(110) single crystal. This is achieved without the need for removal of the well-known Nb(110)-oxide surface reconstruction, due to the introduction of a one-atom-thick Cu interlayer in between the Nb substrate and the magnetic multilayer. The Cu interlayer generates a well-ordered hexagonal surface, which is key for the epitaxial growth of the [Ni/Co]$_{n}$ multilayers hosting the desired out-of-plane anisotropy. The magnetic ground state of the prepared material stacks is directly imaged via spin-polarized low energy electron microscopy (SPLEEM), revealing the presence of magnetic bubble domains with lateral sizes as small as 450 nm. 
\end{abstract}

\maketitle

\section{Introduction}

There is great interest in magnetic materials systems hosting topological spin textures - such as magnetic skyrmions~\cite{WiesendangerNAT-REV-MAT2016,BackJ-PHY-D2020,GobelPHY-REP2021}- due to their great potential for applicability in low-power magnetic data storage devices~\cite{YuNANO-LETT2017} and neuromorphic computing~\cite{HuangNANOTECH2017}. Of particular interest are magnetic thin films, given the ease of combining them with large spin-orbit coupling materials and/or superconductors in heterostructures, which is the desired configuration for solid state technologies.\newline
Currently, there is a fast growing interest in the development of materials platforms in which magnetic skyrmions are placed in direct proximity to a superconducting material. Magnet-superconductor hybrid systems~\cite{LoConteARXIV2024,LinderNAT-PHYS2015} hold great potential for application in quantum computing, since they offer the possibility to host and electrically manipulate topological superconducting states~\cite{HalsPRL2016,BaumardPRB2019,PetrovicPRL2021,RexPRB2019,CollinsSCI-AME2006,BeenakkerARCMP2013,AliceaNAT-PHY2011,NothhelferPRB2022}. This is the concept behind topological quantum computers, which are expected to be more robust against perturbation than current superconducting quantum computers.\newline
In this regard, niobium, Nb, the elemental superconductor with the highest critical temperature ($T_c=9.2$ K) emerges has a natural candidate for this kind of applications, calling for the exploration of the possibility to grow magnetic thin films on Nb single crystal substrates which can potentially host topological spin textures. So far a few attempts have been made in this direction, which focused on the growth of magnetic ultra-thin films on clean and unreconstructed Nb(110) substrates~\cite{LoContePRB2022,BazarnikNAT-COMM2023,GoedeckeACS-NANO2022,SoldiniNAT-PHY2023}. However, those studies required the removal of the commonly present Nb-oxide (Nb-O) surface reconstruction~\cite{OdobeskoPRB2019}, which involves the flash annealing of the Nb(110) crystal at temperatures ($T\approx2400$ $^\circ$C) close to the melting point ($T_{melting}\approx2470$ $^\circ$C). This is a rather tedious and undesired cleaning process, especially if compared to current semiconductor-based technologies' fabrication processes, which calls for the exploration of alternative approaches, where the need for the removal of the Nb-O layer for the growth of magnetic layers hosting the desired spin textures can be circumvented.\newline
Thin films and multilayers hosting magnetic skyrmions are required to possess out-of-plane magnetic anisotropy, given the broken inversion symmetry provided by the magnet/non-magnet interface. A very interesting materials platform in this regard is (111)-oriented [Ni/Co]$_{n}$ multilayers, where perpendicular magnetic anisotropy (PMA) can be stabilized via interface engineering~\cite{AndrieuPRM2018,LoConteNANO-LETT2020}. This makes the combination of [Ni/Co]$_{n}$ multilayers with Nb substrates highly desirable for future quantum technologies. However, so far there have been no reports on the successful growth of highly ordered magnetic thin films and multilayers on Nb(110)-O substrates.\newline
Here we report on the growth of highly ordered (111)-oriented [Ni/Co]$_{n}$ multilayers on Nb(110)-oxide surfaces via interposing an ultra-thin Cu interlayer. The single-atom-thick Cu interlayer induced the formation of a hexagonal surface, on which the growth of (111)-oriented [Ni/Co]$_{n}$ multilayers is enabled. The magnetic ground state of the prepared [Ni/Co]$_{n}$ multilayers is directly imaged via spin-polarized low energy electron microscopy, demonstrating the PMA character of the developed magnetic multilayers and the presence of magnetic bubble domains with lateral sizes as small as $\sim450$ nm.

\begin{figure*}
    \centering
    \includegraphics{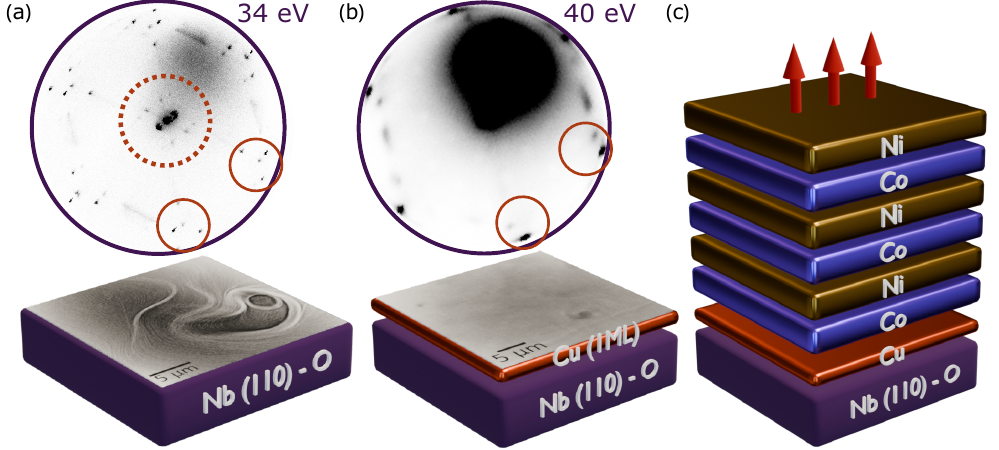}
    \caption{Description of the sample preparation. (a) Schematic of Nb(110)-O substrate with superimposed bright field LEEM image of its surface and  corresponding LEED pattern obtained from the oxidized surface, where some of the satellite peaks, highlighted by a red circle,  correspond to different Nb-O rotational domains. The dotted red circle refers to the LEED pattern shown in Figure \ref{fig:2}.  (b) Schematic of Cu(1 ML)/Nb(110)-O sample with superimposed bright field LEEM image of its surface and corresponding LEED pattern obtained from the sample's surface after deposition of one monolayer of Cu. Satellite peaks are strongly suppressed and an hexagonal patter emerges. (c) Schematic of prepared [Ni-Co]$_3$/Cu(1 ML)/Nb(110)-O sample with out-of-plane magnetization, as indicated by the red arrows.
 \label{fig:1}}
\end{figure*}

\section{Methods}
\subsection{Sample Preparation}
The Nb(110)-O surface is prepared by repeated flash annealing (3 times) up to circa 2000 $^\circ$C in an ultra-high vacuum environment with a base pressure of $5\times10^{-10}$ Torr. The obtained reconstructed Nb(110)-O surface is characterized via low-energy electron diffraction (LEED) and low-energy electron microscopy (LEEM) imaging. The Cu monolayer (ML) is deposited in-situ by physical vapor deposition from an electron beam evaporator in UHV conditions, at a pressure of about $2.0\times10^{-9}$ Torr, while the Nb(110)-O substrate is kept at a temperature of circa 200 $^\circ$C. The Co and Ni layers are deposited on the Cu surface in a similar way, only after the sample has cooled and reached room temperature, at a pressure of about $7-8\times10^{-10}$ Torr.
\subsection{LEEM imaging}
Bright field (BF)-LEEM images are acquired using the back-scattered component of the diffracted electron beam, while dark field (DF)-LEEM images are acquired using the higher order components of the diffracted electron beam. BF-LEEM is used to characterize the morphology of the prepared Nb(110)-O surface, while DF-LEEM is used to characterize its reconstruction rotational domains pattern, respectively.
\subsection{Real-space magnetic imaging}
Real-space magnetic images are acquired via spin-polarized low energy electron microscopy (SPLEEM)\cite{RougemailleEPJAP2010}. In SPLEEM images, the contrast in each pixel is obtained by calculating the asymmetry of the spin-dependent reflection between spin-up and spin-down electron beams, which is $A = (I_\uparrow - I_\downarrow)/(I_\uparrow+I_\downarrow)$. This asymmetry A is proportional to $\textbf{P}\cdot{\textbf{M}}$, where $\textbf{P}$ is the spin polarization vector of the electron beam and $\textbf{M}$
is the magnetization vector. The Cartesian components $M_x$, $M_y$, and $M_z$ of the magnetization are resolved  by  taking  sets  of  images  with  the  electron  beam  spin  polarization  aligned  along  the 
$x$, $y$, and $z$ directions, respectively \cite{RougemailleEPJAP2010}. All images are measured on samples held at room temperature. The energy 
of  the  incident  electron  beam  is  carefully  chosen  for  each  system  in  order  to  optimize  the magnetic contrast.

\section{Results and Discussion}
The sample preparation follows the steps described in Fig. \ref{fig:1}. Figure \ref{fig:1}(a) shows a BF-LEEM image of the prepared Nb(110)-O surface and the corresponding LEED pattern, which reveals the primary bcc(110) spots and additional satellite peaks originating from the different rotational domains of the reconstructed Nb-O surface. Figure \ref{fig:1}(b) shows a BF-LEEM image of the Cu-monolayer deposited on the Nb(110)-O substrate and the corresponding LEED pattern. This LEED pattern is drastically different from the one obtained for the Nb-O surface, showing a strong suppression of the satellite spots and the emergence of a dominant hexagonal pattern. The engineered magnetic heterostructure built on top of the prepared hexagonal surface is shown in Fig.~\ref{fig:1}(c), which is composed of a [Ni/Co]$_{3}$ multilayer hosting a magnetic ground state with the desired perpendicular magnetic anisotropy, as schematically indicated by the red arrows.

\subsection{Oxygen-reconstructed Nb(110) surface}
\begin{figure}
    \centering
    \includegraphics{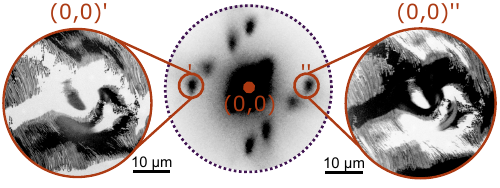}
    \caption{The zoomed-in LEED pattern from the Nb(110)-O surface and two dark field LEEM images, one on the left and one on the right, acquired with the (0,0)' and (0,0)'' satellite LEED beams, respectively. The two dark field LEEM images clearly show the presence of Nb(110)-O rotational domains.\label{fig:2}}
\end{figure}

The details of the prepared Nb(110)-O surface are investigated via DF-LEEM imaging. In order to do so, LEEM imaging is performed selecting the electrons diffracted at specific satellite spots of the LEED pattern, as shown in Fig.~\ref{fig:2}. In the center of Fig.~\ref{fig:2}, the central part of the LEED pattern acquired from the Nb(110)-O surface is shown, with the satellite spots used for DF-LEEM imaging highlighted with red circles and named (0,0)' and (0,0)''. The corresponding DF-LEEM images are shown on the left and the right side of Fig.~\ref{fig:2}. As it is clearly visible from the DF-LEEM images, the Nb(110)-O surface is characterized by the presence of structural rotational domains. In particular, large rotational domains - appearing white and black in the DF-LEEM images - are connected by transition regions consisting of alternating narrow stripes of white and black rotational domains with sub-micrometer width, which appear gray in the DF-LEEM images. All the attempts to grow highly ordered magnetic thin films directly on such reconstructed surface failed, calling for an alternative approach.

\subsection{Cu(111) monolayer on Nb(110)-O surface}
\begin{figure}[h]
    \centering
    \includegraphics{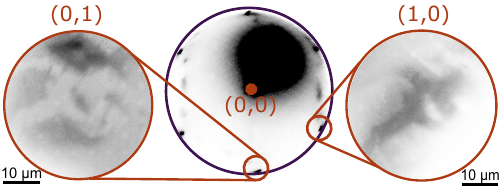}
    \caption{The LEED pattern from the Cu(1 ML)/Nb(110)-O surface and two dark field LEEM images, one on the left and on the right, acquired with the electrons diffracted into the (0,1) and (1,0) LEED beams, respectively. The two dark field LEEM images show a strongly suppressed contrast from the rotational domains present on the Nb(110)-O surface, demonstrating that the Cu monolayer promotes a primarily single domain hexagonal surface.\label{fig:3}}
\end{figure}

With the intent of forming a highly ordered surface with hexagonal symmetry promoting the epitaxial growth of magnetic 3d transition metal layers (e.g. Ni and Co), we developed the growth of a one atom-thick Cu layer. The details of the Cu(1 ML)/Nb(110)-O surface are investigated via LEED and DF-LEEM imaging, as shown in Fig.~\ref{fig:3}. The LEED pattern in the center of Fig.~\ref{fig:3}, which is the same reported in Fig.~\ref{fig:1}(b), demonstrates the presence of a primarily ordered Cu layer, with an hexagonal atomic configuration. DF-LEEM images of the Cu surface are acquired via the diffracted electrons corresponding to the (0,1) and (1,0) spots, as highlighted by the two red circles in Fig.~\ref{fig:3}. The two DF-LEEM images in Fig.~\ref{fig:3} show a strongly attenuated contrast emerging from the rotational domains present in the Nb(110)-O surface. This is evidence of the fact that the Cu monolayer grows in its preferred fcc(111) configuration, regardless of the reconstructed Nb(110)-O substrate underneath. We use this highly ordered Cu(111) surface for the growth of the desired magnetic multilayers.

\begin{figure*}[t]
    \centering
    \includegraphics{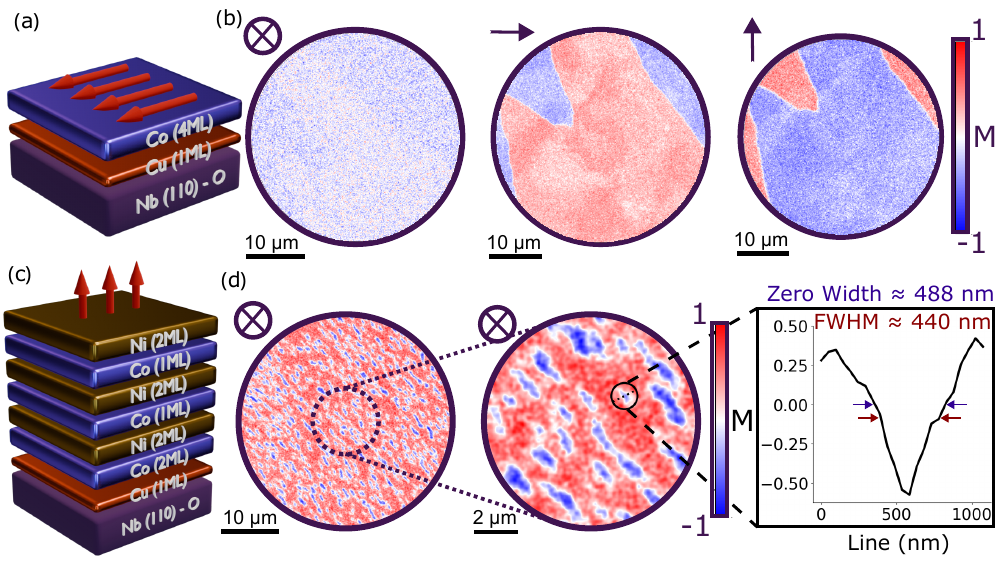}
    \caption{
    Prepared magnetic multilayers and corresponding SPLEEM images. (a) Schematic of Co(4 ML) on the Cu(1 ML)/Nb(110)-O surface. (b) From left to right: out-of-plane, in-plane horizontal and in-plane vertical SPLEEM image of the Co(4 ML)/Cu(1 ML)/Nb(110)-O surface, respectively. The magnetic contrast in the three SPLEEM images is normalized against the same value. (c) Schematic of [Ni/Co]$_3$/Cu(1 ML)/Nb(110)-O with out-of-plane magnetization and layers' thicknesses indicated in monolayer (ML). (d) Out-of-plane SPLEEM image of the [Ni/Co]$_3$/Cu(1 ML)/Nb(110)-O surface, showing the presence of magnetic bubble domains. The zoom-in SPLEEM image in the middle shows the details of the magnetic bubble domains. The graph on the right shows a line-cut across one of the smallest magnetic bubble domains observed in the sample, showing a lateral size of circa 450 nm. \label{fig:4}}
\end{figure*}

\subsection{Engineered [Ni/Co]$_{n}$ multilayer on Cu(1 ML)/Nb(110)-O with PMA}

Next, the highly ordered Cu(111) surface prepared on the Nb(110)-O substrate is used for the epitaxial growth of magnetic thin films of Co and Ni. Figure~\ref{fig:4}(a) schematically shows the first kind of magnetic system prepared, which consists of 4 monolayers of Co hosting an in-plane ferromagnetic ground state, as symbolized by the red arrows. This is proven by the SPLEEM images acquired with electron spin-polarization along three orthogonal directions and shown in Fig.~\ref{fig:4}(b). The SPLEEM image acquired with out-of-plane spin polarization shows no contrast; while the SPLEEM images acquired with in-plane spin polarizations show a strong magnetic contrast. Large magnetic domains with opposite magnetization direction are observed, which is typical of ferromagnetic thin films with in-plane anisotropy. \newline
In order to stabilize PMA in the magnetic system, we took advantage of interface-induced PMA in [Ni/Co]$_{n}$ multilayers~\cite{AndrieuPRM2018,LoConteNANO-LETT2020}. Figure~\ref{fig:4}(c) shows the detailes of the prepared magnetic multilayer hosting a magnetic ground state with PMA. The actual magnetic ground state observed in those layers is shown in Fig.~\ref{fig:4}(d). The SPLEEM image acquired with out-of-plane spin polarization shows the presence of small domains pointing \textit{up}(red) and \textit{down}(blu). More specifically, we observe bubble-like \textit{down} domains, as small as 450 nm in lateral size, in a \textit{up} background. \newline
The observation of magnetic bubbles is a bit surprising, given the fact that such magnetic ground state requires the breaking of time reversal symmetry, usually achieved via an external out-of-plane magnetic field. The most likely explanation for this observation is the small magnetic field produced by the objective lens. Such a magnetic field, pointing perpendicularly to the sample's surface, is expected to be of the order of only 1 mT. This is a very weak field, if compared to the low critical field of superconducting Nb single crystals, which is of the order of 50-75 mT at 4 K. Accordingly, this observation supports the possibility to nucleate magnetic bubbles directly on top of superconducting Nb, and the associated superconducting anti-vortices~\cite{PetrovicPRL2021}, without the suppression of the superconducting state. All this makes the engineered material stack potentially interesting for applications in superconducting spintronics~\cite{LinderNAT-PHYS2015} and quantum computing~\cite{NothhelferPRB2022}.


\section{Conclusion}
In summary, we succeeded in engineering a thin magnetic multilayer above reconstructed Nb(110)-O substrates which hosts a magnetic ground state with out-of-plane anisotropy. This was obtained by growing a [Ni/Co]$_{3}$ multilayer over a one-atom-thick Cu interlayer. The role of the Cu interlayer is crucial for the successful growth of highly ordered epitaxial magnetic multilayers and the stabilization of the desired perpendicular magnetic anisotropy. Finally, the intriguing observations of magnetic bubbles with diameters as small as 450 nm makes the engineered magnetic multilayer interesting for potential applications in superconducting spintronics and quantum technologies.

\section*{Acknowledgments}
R.L.C. acknowledges financial support from the Deutsche Forschungsgemeinschaft (DFG, German Research Foundation) via project no. 512050965. A.D. acknowledges financial support from the Top Master Programme in Nanoscience at the Zernike Institute for Advanced Materials of the University of Groningen. This work was performed at the Molecular Foundry of the Lawrence Berkeley National Laboratory supported by the Office of Science, Office of Basic Energy Sciences, of the U.S. Department of Energy under Contract No. DE-AC02-05CH11231.

\bibliography{jobname}

\end{document}